# AI Exchange Platforms

Johannes Schneider*, Rene Abraham

University of Liechtenstein

Fürst-Franz Josef-Strasse, 9490 Vaduz, Liechtenstein

[johannes.schneider@uni.li](johannes.schneider@uni.li) (Corresponding author*)

rene.abraham@uni.li

The rapid integration of Artificial Intelligence (AI) into organizational technology frameworks has transformed how organizations engage with AI-driven models, influencing both operational performance and strategic innovation. With the advent of foundation models, the importance of structured platforms for AI model exchange has become paramount for organizational efficacy and adaptability. However, a comprehensive framework to categorize and understand these platforms remains underexplored. To address this gap, our taxonomy provides a structured approach to categorize AI exchange platforms, examining key dimensions and characteristics, as well as revealing interesting interaction patterns between public research institutions and organizations: Some platforms leverage peer review as a mechanism for quality control, and provide mechanisms for online testing, deploying, and customization of models. Our paper is beneficial to practitioners seeking to understand challenges and opportunities that arise from AI exchange platforms. For academics, the taxonomy serves as a foundation for further research into the evolution, impact, and best practices associated with AI model sharing and utilization in different contexts. Additionally, our study provides insights into the evolving role of AI in various industries, highlighting the importance of adaptability and innovation in platform design. This paper serves as a critical resource for understanding the dynamic interplay between technology, business models, and user engagement in the rapidly growing domain of AI model exchanges pointing also towards possible future evolution.

## 1 Introduction

In recent years, AI technology has witnessed unprecedented growth, with its applications even permeating domains traditionally associated with "human only" traits, e.g., creativity (Basalla et al., 2022). Recent trends such as generative and agentic AI have further boosted adoption (Schneider, 2025). AI frees engineers from tedious manual and error-prone programming of rules and feature engineering and instead promises to learn from experience in the form of collected data. The product of the learning process is the actual AI, a trained model that can be used for so-called inference, i.e., computing outputs from inputs. Designing and training a model can cost from millions of dollars[1]. Therefore, models can constitute a valuable asset. These assets can be utilized as stand-alone services with minimal additions (e.g., a simple interface as for OpenAI's ChatGPT), or they can be integrated into other products constituting one of many components. For example, in 2019 more than 16,000 apps on Google Play had integrated AI models as a component of an app (Xu et al., 2019). In this paper, we focus on platforms exchanging AI solutions, where we define an AI solution as an "*AI-powered application built around AI models and/or a (pre-trained) AI model.*" That is, the model is of higher relevance than just a (secondary) component. It delivers the core functionality. We consider AI exchange platforms, which we define as *"a digital platform allowing providers to offer AI solutions to consumers"*.

The number of AI models has crossed beyond 400'000 on the largest platform, i.e., HuggingFace. Such dedicated commercial and non-commercial AI exchange platforms for sharing AI solutions reduce search costs (Bakos, 1991), and facilitate access and adoption of AI. Adopting AI often goes well beyond purchasing an AI model. It might come with a multitude of barriers related to financial, personnel, and IT resources, data, culture, etc. (Jöhnk et al., 2021). Platforms handling innovation networks could help in AI-based product innovations (Lyytinen et al., 2016). Such platforms could be highly lucrative as investments in AI are expected to grow from $85.3 billion in 2021, $235 billion in 2024 to more than $630 billion in 2028 (Fang, 2024). Potentially, platforms can differentiate themselves and benefit from offering supportive and common functionalities across a set of AI solutions to facilitate AI adoption, i.e., leveraging economy of scope (Geske et al., 2021). That is, they can go beyond simply offering models through horizontal or vertical expansion (Movahedi et al., 2012). For example, platforms by Amazon, Microsoft, and Google offer machine learning as a service (MLaaS) and AI as a service (AIaaS), which are incorporated into cloud computing platforms (Lins et al., 2021; Yao et al., 2017). However, as of now, it is unclear to what extent and how AI exchange platforms differ and there is a need to better understand how platforms support leveraging the core of AI: AI models, especially, since there have been multiple (potentially harmful) incidents on common platforms (Gorwa & Veale, 2023).

---

[1] https://www.forbes.com/sites/craigsmith/2023/09/08/what-large-models-cost-you--there-is-no-free-ai-lunch/

Furthermore, AI is a transformative technology that is likely to disrupt many industries. However, it is highly complex and notoriously difficult to understand. At the same time, there is not just one AI model, but AI models can be trained on different data (e.g., text, images, audio), different tasks (e.g., classification, generation), and can be tailored to specific hardware (e.g., large scale servers or edge devices). Furthermore, such models operate under very different legal frameworks, e.g., image classification for autonomous cars, medical devices, and ordinary smartphone camera apps are all subject to different regulations. Thus, while digital platforms are known for the winner-takes-all effects, it is unclear whether there is not any room for more specific platforms. Additionally, AI is quickly evolving. Recently, adoption and interaction with AI have become much easier for end users with the rise of foundation models (Bommasani et al., 2021; Schneider et al., 2024) that can follow natural language instructions (so-called prompts) to accomplish a wide range of tasks. But even leaders in the era of foundation models such as OpenAI have recognized the value of providing a platform where users can offer and consume custom versions of their foundation models or build plugins around their foundation model - similarly to large software vendors aiming at "inverting their firm using external developers" (Parker et al., 2017). However, AI models and software differ in many aspects, e.g., AI is a black box surrounded by a large ethical debate and rapidly evolving legislation that might impact platforms seeking to facilitate access to AI models. Therefore, there are likely significant differences between conventional developer platforms such as "App stores" and AI exchange platforms that we set forth to uncover. Furthermore, the "No code" and "low code" (Bock & Frank, 2021) paradigm offers increasingly more possibilities to produce AI solutions at lower costs and with less expertise. For example, OpenAI's GPT store allows ordinary users to customize (one of) the most powerful large language models through a web interface by simply stating prompts and instructions combined with data uploads. It enables ordinary people to monetize their adaptations of ChatGPT directly on the platform. Given the surge in the number of AI solutions over the last years and the fact that the number is likely to increase, the demand for market coordination services like AI exchange platforms is also growing.

Looking at the existing research in the literature (e.g., Berente et al., 2021; Lyytinen et al., 2016; Van den Broek et al., 2021), we observe a general orientation toward organizational and managerial implications, but the business side of AI remains obscure. Taxonomies help practitioners and researchers understand complex domains (Nickerson et al., 2013). They can constitute the first step in rigorous theory building (Williams et al., 2008). This leads to our research question: *Which dimensions, characteristics, and archetypes allow describing existing AI exchange platforms?*

We address the question by building a taxonomy based on Nickerson et al. (2013) and identifying archetypes using cross-case analysis (Yin, 2009). We combined empirical knowledge found in 23 platforms with conceptual understanding found in academic literature, taking the lens of producers and consumers of AI solutions. We contribute by providing a structured overview and systematization of AI exchange platforms in the form of a taxonomy and archetypes. Our discussion also surfaces interesting observations and avenues for future research related to research policies, the evolution of platforms, and their designs.

## 2  Research Background and Related Work

Our work essentially relates to two important areas: AI, more specifically AI models, and digital platforms, more specifically multi-sided platforms. An AI model is a software that takes an input and produces an output "without explicitly being programmed", as coined by Arthur Jackson in 1959. That is, it originates from a training process based on (historical) data. Pre-trained AI models can be further customized to other applications by fine-tuning based on problem-specific data, commonly held by a company wishing to adjust AI to its needs. This process is also known as transfer learning (Zhuang et al., 2020). While pre-training a raw, untrained model involves commonly large amounts of data and computing (Bommasani et al., 2021), fine-tuning might be performed with much less data and computational effort. For example, using a technique called QLora large language models(LLMs) such as ChatGPT can be fine-tuned on a few GPUs within less than a week (Dettmers et al., 2023). Thus, it becomes feasible for any business to leverage and customize LLMs to their specific needs. LLMs fall under the category of foundation models (Bommasani et al., 2021) that can be used for a variety of tasks and have become more and more powerful. Still, according to the free lunch theorem by Wolpert & Macready (1997), no single model can be best for all tasks. Thus, there is an inherent need for multiple models to address different tasks. Furthermore, even for foundation models (like OpenAI's GPT series), task-specific performance can be enhanced with fine-tuning or prompt engineering both more cost-effective than developing a new model (Liu et al. 2023; Niu et al. 2020). The diversity of models is further increased as they can be defined using a specific development framework (such as Tensorflow or PyTorch). In summary, there exists a large amount of AI models that can be adjusted to specific needs and, in turn, call for AI exchange platforms for support.

AI models also exhibit special traits that distinguish them from other technologies and, in turn, could potentially impact exchange platforms. AI models are difficult to understand (Meske et al., 2022). They can act unethically (Grewal et al. 2021), e.g., deceptively (Schneider et al. 2022b), and they can face privacy (Truex et al., 2019) and security concerns (Chakraborty et al., 2018). It is difficult to assess if a model performs well in slightly different contexts than the scenarios it was trained for, as well as if it behaves ethically and legally compliantly despite extensive research on how to explain models (Meske et al., 2022). These aspects pose challenges for providers that constitute opportunities for AI exchange platforms to generate value by offering (additional) services.

Given the dynamics of AI and its complexity, any digital platform helping in the organization of the AI market is likely of value. Under the conceptualization by De Reuver et al. (2018), we investigate multi-sided platforms and multi-sided markets. Our definition of an AI exchange platform is more economically inspired; however, our analysis also covers technical aspects, i.e., we account for both engineering and economic perspectives of digital platforms (Gawer, 2014). In our work, investigating platform boundaries (Gawer, 2021) is also of interest to gain a more holistic view of AI exchange platforms, i.e., the platform scope (what assets are owned, what activities are performed) and the configuration and composition of platform sides (customer groups).

There are a variety of works that tackle taxonomy building. Freichel et al. (2021) developed a taxonomy of digital platforms based on a literature review, which is generic and thus misses specific aspects related to AI. Blaschke et al. (2019) employed an architectural perspective to build a taxonomy of digital platforms. In light of their taxonomy distinguishing service, ecosystem, core, and infrastructure as dimensions and their archetypes, we observe that AI exchange platforms do not fit any of the listed archetypes. Furthermore, AI exchange platforms cover multiple of their listed characteristics (of a dimension), hinting that AI exchange platforms cover a broad range of digital platforms. Geske et al. (2021) developed a taxonomy of AI service platforms based on a sample of 31 platforms. We focus more on the AI (model) "itself" that solves a problem, rather than developer tools or infrastructure. That is, their view is much broader and only four platforms intersect with our work. They include platforms that are completely independent of models such as platforms for data labeling. Lins et al. (2021) distinguish three levels for AIaaS depending on the capability provided, i.e., ready-to-use services, developer tools, and infrastructure services, which are also mentioned by Geske et al. (2021). We believe that putting the actual product, i.e., AI models, in the center is a valuable additional perspective as there is limited overlap with prior work - not least because models become more and more powerful with the rise of foundation models (Bommasani et al., 2021). Few works have focused on AI model "stores" and AI marketplaces. Xiu et al. (2020) compared three model "stores" with two app stores. Differences were found in the information stated on the product and in the documentation (e.g., release notes, version number, training set) and deployment, i.e., the deployment was in the cloud for models and on the device for apps, and user feedback was consistently allowed on all model stores. However, the sample size was small containing only three model stores. Kumar et al. (2021) sketch AI marketplaces' technical, economic, and regulatory aspects. They discuss multiple stakeholders of an AI marketplace including auditors and regulators. Their conceptual findings are sketched as stated in the paper's title, i.e., they are not based on an empirical investigation. They also consider data marketplaces and interoperability standards as marketplaces. Out of the 24 listed marketplaces only two intersect with our identified platforms. Multiple works also proposed technical implementations of AI marketplaces (Sarpatwar et al. 2019, Somy et al. 2019). Such marketplaces might employ AI themselves to facilitate interaction. According to Rai et al. (2019) such a platform would be an AI platform being defined as a "next-generation digital platform, arising from the application of artificial intelligence (AI) technologies." Data and model markets have also been introduced in the form of a tutorial to the database and data management community (Pei et al., 2023). In contrast, we do not focus on data markets and we do not elaborate in depth on technical implementations of such platforms. Most recently, (Qian et al., 2025) surveyed model marketplaces focusing primarily on operational aspects. They include companies supporting model access for development (like DataStax), which we do not. Also they include companies that list (their) use cases rather than actual products and do not allow providers to upload their solutions (like Infotech "marketplace"). The paper contains 14 marketplaces, while we include more than 20. Our work develops a taxonomy including platform archetypes and dimensions. It is also less technical.

## 3 Methodology

A key problem in many disciplines is to derive conceptual knowledge and, more specifically, to classify objects in a domain of interest into a taxonomy. A taxonomy "is a set of dimensions consisting of mutually exclusive and collectively exhaustive characteristics" (Nickerson et al., 2013). To develop the taxonomy, we used the iterative approach by Nickerson et al. (2013) synthesizing both existing conceptual knowledge from literature as well as empirical knowledge from analyzing objects of interest.

The process asks for an initial definition of meta-characteristics and ending conditions. Then, it iteratively identifies new dimensions and characteristics to revise the taxonomy until the ending conditions are met. The identification is done either based on an empirical-to-conceptual (e2c) approach or vice versa (c2e). The meta-characteristic is the most comprehensive characteristic that others should follow. We decided to apply the lens of business models, which is aligned with our more economically oriented definition of AI exchange platforms. More precisely, we used the three common elements of value creation, value delivery, and value capture (Remane et al., 2017). Our work also includes the discussion of non-commercial platforms for which the "value capture" meta-characteristic is omitted since it relates to the profitability of platforms. As ending conditions we used the ones in the original work (Nickerson et al., 2013) distinguishing between subjective conditions such as conciseness, robustness, comprehensiveness of the taxonomy, and objective conditions such as that taxonomy was not altered in the last iteration, it is free from duplicates (both in terms of dimensions and characteristics of a dimension), and all objects were investigated.

We performed a total of nine iterations to obtain the taxonomy. In the first iteration, we empirically assessed existing marketplaces to derive a first set of dimensions and gain an initial overview of the possible characteristics of each dimension. One of the authors, who has used a number of models from different frameworks and AI exchange platforms suggested searching for "model marketplaces", "AI marketplaces", "AI model exchange platform" and "model zoo", which is a collection of pre-trained models. We also assessed if there are other relevant terms in related work. We searched on different dates around the 1st of March 2022. We performed a second search with additional terms prior to iteration seven around the 1st of January 2024. We performed a third round in August 2025.[2] We investigated at least the first 50 hits on Google and stopped once 30 hits in a row did not yield any novel platforms. We also took platforms into account that were mentioned in other works (Geske et al., 2021; Kumar et al., 2021; Lins et al., 2021). However, this did not result in any additions since we covered most platforms in these works. The ones present in other studies (but not in ours) were often no longer operational or did not meet inclusion criteria, e.g., they were data marketplaces not allowing the exchange of AI solutions. We required that platforms were operational, offering at least ten AI solutions, and showed some activity in the last 1.5 years prior to the search. Platforms that were shut down at the time of search[3], superseded by another platform (such as the Caffe2 model zoo), or under development were not included. Furthermore, we excluded platforms that sold software where AI models were not central, i.e., AI was a (small) component or did not explicitly categorize offerings as AI(-based) (e.g., Google app store) or did not allow to explore AI solutions (e.g., Nuance). That is, while we allowed that an application was only built around an AI model (e.g., as in OpenAI's plugin store), it was essential that the core functionality of the application stemmed from the AI model and the AI exchange platform was also clearly listing AI solutions separately from non-AI based applications, e.g., the Unity platform contains a category for AI applications and within this category an application that contains speech recognition, which relies on a text-to-speech AI model. A platform called "Blackford" contained medical image applications that relied mostly on an AI model. Both exemplary apps were accompanied by a user interface and other functionality, but the essential functionality stemmed from the AI model. In case, it was unclear, if an AI model was really central to the offered artifact, we did not include it in our "AI solution" count. Counts were often a coarse estimate as counts were not always listed on the website and it was impossible to manually check in the order of millions offerings. To assess platforms, we took the role of a consumer and a seller. We registered online to access a platform, i.e., to browse AI solutions. We did not contact sales representatives, perform payments to access AI solutions, or inquire about platform-specific contribution requirements if they were not explicitly stated on the platform. As a consumer, we browsed AI solutions, chose a few, read through model descriptions, tried to compare them based on provided information, and tried to access AI solutions. We also took the role of a contributor, assessing how to upload an AI solution, what to consider in terms of technology to build an AI solution, required documentation, contractual obligations, and compensation schemes for offerings. Included platforms also had to explicitly encourage contributors to submit AI solutions of their choice, e.g., a few platforms asked only for the implementation of specific models described in a given list of research papers. Out of more than 94 initial platforms, we were left with 30 models for the first iteration and 26 in the final selection (see Table 1).

---

[2] We submitted this paper to reputable journal, but it stayed with editor for 11 months despite multiple messages. So we had to withdraw and decided that we needed another round of search.

[3] If the platform was shutdown between two searches, we kept it but marked it as shutdown. This was only true for one platform.

| Platform | URL | #AI solutions |
| --- | --- | --- |
| HuggingFace | https://HuggingFace.co/ | ~400'000 |
| **Civitai** | https://civitai.com/models | ~230'000 |
| OpenAI's GPTs | https://chatgpt.com/gpts | ~150'000 |
| PapersWithCode* | https://PapersWithCode.com/ | ~70'000 |
| Kaggle | https://www.kaggle.com/models?tfhub-redirect=true | ~3'400 |
| Amazon Sagemaker Marketplace | https://aws.amazon.com/marketplace/search | ~3'000 |
| Azure | https://azuremarketplace.microsoft.com/en-au/marketplace/apps | ~2'000 |
| Google AI | https://cloud.google.com/marketplace | ~2'000 |
| **Databricks Marketplace** | https://marketplace.databricks.com/ | ~1'100 |
| **Modelscope** | https://www.modelscope.ai/ | ~1'100 |
| OpenAI's Plugin Store* | https://chat.openai.com/?model=gpt-4-plugins | 1023 |
| ModelZoo.co | https://modelzoo.co/ | ~1'000 |
| Pinto | https://github.com/PINTO0309/PINTO_model_zoo | 430 |
| OpenVino Model zoo | https://docs.openvino.ai/nightly/omz_models_group_public.html | ~250 |
| Gravity AI | https://www.gravity-ai.com/ | ~150 |
| ParlAI | https://github.com/facebookresearch/ParlAI | 130 |
| Tensorflow models / Model garden | https://github.com/tensorflow/models | ~100 |
| Unity Assetstore | https://assetstore.unity.com/tools/ai-ml-integration | ~100 |
| Sectra Amplifier Marketplace | https://amplifiermarketplace.sectra.com/ | 85 |
| Singularity | https://beta.singularitynet.io/ | 84 |

| Pytorch Model zoo | https://pytorch.org/hub/ | 52 |
| Workday | https://marketplace.workday.com/ | 35 |
| AI Tech | https://ai.aitech.io/ | 33 |
| AI Planet | https://app.aimarketplace.co/ | ~20 |
| Blackford | https://www.blackfordanalysis.com/blackford-dashboard/ | ~20 |
| Neuromarket | https://neuromarket.ai/ | 13 |

Table 1: Taxonomy of AI exchange platforms (* terminated)

For each platform, we collected two sources of evidence. The first stemmed from general descriptions found on its website, e.g., descriptions of the platforms, offered services, white papers, and other official publications. The second stemmed from interaction with the platform, i.e., when we took the role of a consumer and a provider.

The second iteration was conceptual-to-empirical. We assessed the existing literature on taxonomies of platform business models and platforms (Täuscher & Laudien, 2018). We also reviewed literature on AI services, AI exchange platforms, and data marketplaces, as presented in the related work section. In particular, we investigated technical proposals to build such platforms as a source for possible dimensions and characteristics, e.g., how to build trustless and ownership-preserving marketplaces (Kurtulmus & Daniel, 2018; Somy et al., 2019).

Iteration three to six were smaller in terms of changes and additions to the framework. They consisted of refinements of the framework based on reflection and related work. They alternated between e2c and c2e. For example, we summarized and generalized some characteristics based on their frequency. Initially, we tracked the metrics of models in more detail (such as GFlops and accuracy). Here, we grouped them into the categories performance metrics and computation needs. Iteration four was c2e, i.e., we expanded on related work of our empirically identified concepts. Iterations five and six were both e2c and yielded only small and no changes, respectively.

Prior to iteration seven, which was conducted more than a year after the prior iterations, we performed another search for platforms as described for iteration one but only included pages from 2022 onwards. We also expanded the search terms including "AI exchange platform" and "AI plugin store".[4] We also reviewed all prior AI exchange platforms, and whether they still meet the inclusion criteria. This led to the inclusion of nine new platforms and the exclusion of 13 platforms as they showed no activity in the last 1.5 years. Most of these platforms were still accessible though. Some platforms also migrated, e.g., Google's Tensorflow Hub was redirected to the Kaggle platform. We performed an e2c, which left the dimensions unchanged but we altered characteristics. We improved the nomenclature and removed and added characteristics, e.g., we removed "freemium" as only one marketplace offered it and added "indirect access (through a vendor)". The nineth iteration was also an update – identical to the description of the prior paragraph (see also Footnote 2 for reasons why this was necessary). We added six more platforms (among them three from (Qian et al., 2025)). We also checked the others, which did not yield further changes (the shutdown time of the OpenAI pluginstore was less than 1.5 years).

---

[4] "AI plugin store" was added to capture plugin stores such as the one by OpenAI. AI marketplace was added as prior work focusing on AI marketplaces as prior work searching for AI marketplaces (which we also used to expand our list of platforms) has become outdated. We added "AI exchange platforms" as we expanded our scope beyond "AI model exchange".

In our research, we also identified archetypes by performing a cross-case synthesis (Yin, 2009). Cross-case synthesis retains (most of) a case and compares within-case patterns across the set of available cases. Thus, we analyzed each case in its entirety and conducted pair-wise comparisons looking for similarities and dissimilarities. This eventually allowed us to identify groups of related cases based on their characteristics.

# 4  Results

We first provide our taxonomy followed by four archetypes.

## *4.1  Taxonomy of AI exchange platforms*

An overview of the taxonomy including all dimensions and characteristics for each dimension is given in Table 2. As meta-characteristics, we employed fundamental elements of business models, i.e., value creation, delivery, and capture (Täuscher & Laudien, 2018).

**Value creation**

*Infrastructure:* The infrastructure dimension describes the design of the multi-sided platform focusing on where AI solutions are hosted prior to and during transactions. In a decentralized platform, they reside outside (the control of the platform), e.g., they might reside within a blockchain, at a private server of an app provider, or at a public repository hosting code (and other digital artifacts), e.g., GitHub and Google Drive. In a platform with centralized storage, AI solutions are stored as part of the platform prior to a transaction. A consumer might often download the AI solution prior to its use. Multiple platforms are coupled with cloud computing infrastructure allowing to run or host the AI solution within a cloud infrastructure without any additional actions by the AI solution providers and consumers. Thus, neither of them requires additional infrastructure to operate AI. For "hosted" AI solutions, consumers could either send inputs to a server managed by the platform, potentially handling requests from many customers or rent dedicated, i.e., non-shared, computational resources on which they could deploy and potentially alter an AI model in an easy manner. For example, Amazon Web Services (AWS) allows renting computational infrastructure on their cloud. The two medical-focused platforms did not state where hosting takes place.

| Meta-dim. | Dimension | Characteristics | | | |
|---|---|---|---|---|---|
| Value creation | Infrastructure | Decentralized (4) | Centralized storage(7) | Centralized storage + execution(13) | Not stated (2) |
| | Online testing | Yes(8) | No(18) | | |
| | Review | Internal Only(16) | User(7) | Academic Peer review(3) | |
| | Quantitative comparison | Performance metrics (2) | Usage/Rating (9) | None(15) | |
| | Contributor attraction | Little effort(4) | Free services(2) | Financial(15) | None(5) |
| | Data | No(10) | No, but information on training data (7) | Yes, datasets available (9) | |
| | Training procedure | No info(17) | Source code available(2) | Only paper available (7) | |
| | Customization | Yes, but not on the platform(8) | Yes, offered as a service(5) | No (13) | |
| Value delivery | Access | Indirect (through vendor, not platform) (3) | Download only(9) | API/Web only(6) | Download +API(8) |
| | Domain | Agnostic(24) | Specific(2) | | |
| | AI branch | All(19) | Vision(3) | NLP(4) | |
| | Target environment | All(11) | Specific(15) | | |
| | Participants | B2B(2) | All2All(12) | All2B/All with research emphasis(11) | B2All(1) |
| Value capture | Revenue source | Sales of product (8) | Sales of other services(7) | Platform only complementary (7) | None(4) |
| | Participants charge | One-time fee(3) | One-time fee and/or Usage(13) | None(8) | Not stated(2) |

Table 2: Taxonomy of AI exchange platforms

*Online testing:* Testing refers to obtaining model inferences for user-provided data. Online testing implies testing without the need to perform any significant deployment effort, in particular without any local installation, e.g., by uploading data directly on the website of the platform and obtaining outputs. For platforms billing customers per inference, testing a few instances was typically cheap and sometimes included as part of a freemium model. OpenAI's GPT store also effectively provides online testing, as the execution of a customized model by a provider on the GPT store, could be done within a few seconds. Offline testing requires often significantly more effort to get access to a model.

*Review* is the process of assessing an AI solution either prior to listing it on the platform, e.g., to reject it before consumers are aware of it, or after it has been made available on the platform, e.g., allowing users to share their experiences with the solution using reviews. Reviews might be seen as a form of quality assessment that can act as a preventive measure to avoid low-quality AI solutions entering the market in the first place or as a reactive measure ensuring that low-quality solutions are labeled as such. Marketplaces can also foster trust by providing reviews of participants' prior transactions (Pavlou & Dimoka, 2006). Platforms like Amazon allow users to review AI solutions like any other purchasable product on the marketplace. All platforms perform some form of curation. No platform allowed to directly upload artifacts without any form of registration or granting permission.

Most internal reviews were done according to predominantly non-specified procedures, e.g., some platforms dedicated to models asked a contributor to get in touch before submitting a model. Most details on the review process could be found for OpenAI's plugin and GPT store, which included compliance with their content policy (e.g., excluding harmful applications) and brand guidelines, enforced quality constraints, and the provision of metainformation. The plugin store also demanded human agency of actions, i.e., users must confirm any action by the plugin before it could be taken. This is aligned with the European AI Act that demands that AI applications require human oversight (depending on their risk categorization). Platforms themselves might also review providers (rather than just AI solutions), e.g., they might engage in "partnerships" with AI solution providers given that they match specific criteria (which none of the platforms publicly disclosed in detail). For example, small, specialized platforms from the medical domain (like Sectra), required providers and consumers to reach out to the company (personally) rather than transacting online.

Overall, the review systems are highly interesting since some contain characteristics not commonly found in other digital platforms. That is, some platforms rely on external reviews. External reviews can stem from institutions such as the US Food and Drug Administration (or its European counterpart) for models in the medical domain, but also academic peer review. For some platforms, it was sufficient if model-related papers were non-peer-reviewed papers, e.g., published on arxiv.org. Some platforms even automatically added papers and code repositories. Other platforms explicitly stated that models must be described in papers published in reputable conferences or journals. Thus, they trust in academic peer review and effectively benefit from it. For example, contributing a model to TensorFlow requires that models be published at a top-tier machine learning conference. While some platforms did not restrict themselves to academic model contributions, others effectively only contained models with references pointing to research papers. Intel's OpenVino (and Amazon) also required that models pass basic tests, e.g., that they could process inputs. Other platforms do not have explicit reviews but provide some indication of the popularity of AI solutions, e.g., download statistics over time and likes (e.g., HuggingFace) or a non-defined popularity ranking (e.g., OpenAI's plugin store).

*Quantitative comparison:* This dimension describes how consumers can rank AI solutions, i.e., models, based on common quantitative measures capturing important model properties, usage information, or quantitative user ratings. Thus, qualitative reviews, e.g., textual comments, are not included. Options for model comparison were mostly limited or not existing, i.e., although many models were made for the same task and possibly even trained and evaluated on the same dataset with outcomes stated in the paper. Some platforms applied holistic measures correlating with overall customer satisfaction, such as reactive measures, e.g., likes could be found and used for comparison. Some platforms also provided usage statistics. Few marketplaces offered to compare models based on typical assessment metrics. They provided computational metrics, such as required computation [Gflops], inferences per second, model size (number of parameters), and performance metrics, e.g., accuracy for models evaluated on the same dataset.

*Contributor attraction:* This dimension describes what makes a platform attractive for contributors. To investigate why some platforms might be more attractive to contribute to than others, we looked for features or explicit statements incentivizing or facilitating contributions to the platform. Though only two platforms explicitly stated visibility as a reason to contribute, the same argument can be made for any platform and, therefore, was not listed. Visibility is essential for researchers and commercial vendors alike. Scientists are often assessed based on their citations, which are likely to increase if their models are used. Platforms also commonly listed AI solutions with references to the original articles describing the underlying model. Platforms like PapersWithCode (and to smaller degree Huggingface and others) provided explicit leaderboards for various benchmark datasets common among researchers for model evaluation. Being on top of a well-reputable leaderboard might be appealing for researchers since AI research is typically very competitive, i.e., there is a constant race to outperform existing AI approaches trained on common public benchmark datasets. The HuggingFace platform explicitly stated the benefits of joining as getting easy access to a large set of resources (other models and datasets), free project management, e.g., versioning and branching, and better discoverability. The effort to add AI solutions varied significantly by platform. Some platforms required very little effort, e.g., submitting only a link to a public code repository such as GitHub, which many researchers use anyway since conferences and journals encourage code submissions typically hosted on such platforms. As such, submitting a link is little effort compared to other platforms that often require manually declaring metadata, adhering to specific interfaces, running tests, etc.

*Data:* The data dimension relates to the availability of data on the platform, i.e., transaction content. Many platforms offered datasets in addition to AI solutions. These datasets could be acquired independently of an AI solution. That is, multiple platforms allowed the exchange of both datasets and AI solutions. A special case was Civitai, where data came in the form of single samples that were typically AI-generated images rather than entire datasets. For a consumer, it is often of interest to get the training data used to train a particular model. Platforms that only offered ready-to-use models or AI-powered applications were mostly non-transparent, i.e., they mostly did not state any information on the datasets and the model architecture. Some platforms provided links to a model's (training) data as part of the model description or at least referenced the paper, allowing us to determine the data used. However, even for academic works that strive for transparency and reproducibility, access to training data was often neither available through the platform nor by reading the paper and searching manually for the dataset, since datasets were non-public.

*Training procedure:* This dimension relates to transaction content. It describes whether AI solutions were accompanied by information on how models were trained, such as details on the employed model optimization technique and data augmentation techniques. Platforms offering only ready-to-use models did not provide any information. A few platforms provided direct links to source code though often only for some models. Commonly, offered AI solutions for download were accompanied by references to academic works that stated information on the training procedure, but it was not always clear, whether the models followed these procedures.

*Customization* describes whether the platforms offer AI solutions that consumers can further adjust. If so, we also distinguished between platforms that offer customization as a service. In addition, some platforms offered data or sold developer tools supporting customization. Commonly, the transaction content is a service without customization. That is, ready-to-use models are used for inferences, i.e., the model is accessed by sending inputs and receiving outputs, where the platform might host the model. If not, there are various means for customization demanding a different level of skillsets and data. Downloadable AI models provided in standardized formats or for well-known frameworks can be adjusted by the users to their own needs. That is, data scientists or machine learning engineers can fine-tune them for a specific task based on a proprietary dataset possibly on their computing platform. This process can be supported to different degrees by services on the platform. For example, the websites of marketplaces of the three large cloud providers (Amazon, Google, Azure) often had a category for ML services that would include both models and developer tools offered on the marketplace. Such services might take care of the actual customization process and only require uploading (training) data, usually in a well-specified format. While this simplifies development and reduces the needed AI expertise, it still requires some expertise. When model customization was offered as a service, it sometimes included model selection, i.e., a user only provided a dataset (and not a pre-trained model), and the platform automatically identified a suitable model. The recently launched GPT-store by OpenAI (superseding its plugin store) goes past many prior customization service offerings by allowing to adjust large language models, i.e., ChatGPT, based on textual instructions ad (optionally) with arbitrary (non-public) data. The customized LLM can also be tested on the platform. Thus, customization can essentially be done by laymen.

However, usage rights might restrict modifications of AI solutions. For commercial vendors of ready-to-use models for a specific application, we found examples where models were equated to a piece of software that should neither be altered nor be reverse-engineered (according to the EULA). On the other hand, on academic platforms, such as PapersWithCode, a customizable model often consists of a pre-trained model, source code facilitating model fine-tuning, references to the training data, and detailed documentation on the model and its architecture, i.e., in the form of a paper. Furthermore, licenses mostly did not restrict customization.

**Value delivery**

*Access:* Access refers to how consumers can obtain the transaction content. Models can be accessed in different ways. Most platforms allowed model download.[5] Two platforms only held links to repositories containing the models from which they could be downloaded. API access or access through a web interface, for hosted models was supported by eight platforms. AI solutions providers had to upload their AI solutions accompanied with information on inputs and outputs or adhere to specific guidelines on how to offer services, which posits additional work. For example, SingularityNET requires model providers to adhere to a specific design, e.g., a model must be run in a service including a daemon that can interact with the platform. The daemon obtains requests from platform users, checks funds, and forwards the requests to the model.

*Domains:* Domains refer to the economic sectors that a platform targets. Two platforms focused only on one particular economic domain, i.e., medical. The remaining platforms were not limited to a particular domain. Though some listed a number of industries. Overall platforms could also have been distinguished based on pure model platforms, pure application-focused platforms (having the model at a core) and those allowing both.

---

[5] We also said that a model was downloadable, if it could be deployed from the platform in a dedicated virtual instance for a consumer, and from there it could be downloaded.

*AI branch:* AI branch refers to the subfield of AI for which the platform offers models. Most platforms were not restricted to any AI branch, i.e., all but three supported diverse tasks from areas such as speech processing, natural language processing (NLP), and computer vision. Though offering models from various task areas, some platforms were somewhat specialized in one AI branch. For example, HuggingFace offers diverse models but focuses on NLP historically. Specialized platforms focused on computer vision or NLP. The GPT store allowed to access vision and NLP models (Dall-E and ChatGPT).

*Model target environment:* This dimension states the environment, including hardware and software inference engines, for which the offered models are developed or optimized, e.g., standard development frameworks (e.g. PyTorch and Tensorflow) or proprietary environments (OpenAI). Development frameworks allow the specification of an entire development pipeline including data preprocessing, model training, tuning, and evaluation. While there are efforts to develop open standards to port ready-to-use models, e.g., ONNX, the entire pipeline is not easily portable. Models can also be optimized toward specific hardware to reduce inference and training times as shown by Intel's OpenVino platform. Models that are optimized for specific hardware might rely on hardware vendor-specific features and, therefore might not work on other hardware.

*Participants:* This dimension describes the parties involved in transactions as contributors and consumers. For most platforms, due to the description of the platform and the offered models, we identified two participants involved in transactions on the platforms: researchers (or hobbyists) and businesses. Still, access to most platforms was unrestricted, allowing any person to access models. However, end consumers are more likely to use AI that comes in applications abstracting from the need to install and utilize special frameworks to run models or get acquainted with web services to use APIs. Still, there are indications (such as discussions on Reddit, and blog posts on Medium) that hobbyists are using such models. Also, the new GPT store encourages any user of ChatGPT (e.g., beyond 100 million people worldwide) to create and offer their own AI solution on their platform. On the contributor side, while AI-powered solutions leveraging AI models are developed by a large group of people, we did not find any (innovative) AI models that stemmed neither from an organization nor were linked to an academic publication, counting also papers on preprint-repositories, i.e., arxiv.org. That is, (re)implementations of papers, e.g., for different frameworks, and adjustments of models were often carried out by hobbyists, and were also frequently used by consumers. For example, Civitai contains a large number of text-to-image models that were fine-tuned using (open-source) models by hobbyists for a large number of different image types. For platforms that were open to everyone, we used the term "All" indicating that businesses, (ordinary) consumers, and researchers are potential participants.

Researchers could easily use some commercial platforms to sell their products, e.g., Amazon offers an "individual seller account". Opening such an account requires some administrative work. Among several hundred investigated offerings, we found none that were contributed by researchers in the role of an individual, e.g., not being part of a company such as a start-up, but only by companies. For example, a company called "Extrapolations" packages well-known, state-of-the-art models from research papers from various authors for various tasks and offers them on Amazon. Amazon Web Services also offers some standard, commonly used models from different technical platforms, e.g., a ResNet network for image recognition developed using the Pytorch and MXNet platforms.

Some platforms require that models must be based on a scientific publication (or at least a preprint on arXiv, which requires an academic mail) though the contributor itself does not necessarily have to be a researcher, e.g., PapersWithCode states "Anyone can contribute - look for the 'Edit' buttons! Want to submit a new code implementation? Search for the paper title, and then add the implementation on the paper page".[6] To contribute to Tensorflow (as part of the research branch), the requirements for model selection are: "A model from the paper accepted at top machine learning venues" or "A state-of-the-art model from a pre-publication available at arXiv".[7] Platforms that explicitly demanded such criteria, required a reference of a paper (e.g., OpenVino) or explicitly stated dissemination/support of research (e.g. ParlAI) were classified as "All2All/B" with research emphasis". Some platforms did not explicitly target researchers as contributors or users, i.e., they did not state that a scientific publication or preprint was necessary, though all available models stemmed from scientific publications with accompanying references. These platforms were also not transparent about their internal review processes for model contributions. Thus, it might be assumed that a scientific publication is at least a facilitator for acceptance. We classified the contributing participants for these platforms as All2All. For example, HuggingFace claims to have more than 5,000 organizations including universities, industrial research facilities, and regular companies as contributors. A few platforms were also primarily targeted toward researchers according to their websites' descriptions but did not restrict access. For example, there exists a platform for "dialog research" (e.g. ParlAI). We classified them along with other platforms that invited researchers and companies and, more generally, any open-source contributors, also as All2All. Even the medical focused platform by Sectra had "research only" models with references to publications, while the other medical focused platform, i.e., Blackford, was the only platform that was only B2B.

---

[6] https://paperswithcode.com/about

[7] https://github.com/Tensorflow/models/wiki/Research-paper-code-contribution

**Value capture**

*Revenue source:* The revenue source describes how platform owners create income by billing either sellers or consumers. Many platforms were non-commercial and did not generate any revenue. They were mostly initiated and funded by public or industrial research institutions. Some platforms are complementary to existing services or products. These were operated by corporations that typically only offered models for a particular development platform or optimized them to execute on specific hardware or software environment or offered non-AI products allowing to integration of AI functionality (e.g., Sectra for medical imaging, Unity for speech recognition, etc.). Thus, the platform might benefit a company's entire product ecosystem. The Intel corporation offered optimized models for their hardware products, i.e., processors. Finally, six platforms offered additional services to charge users, such as hosting models in the cloud and obtaining the best model for their problem (or dataset), i.e., MLaaS or implementation consulting. A few platforms offered models for free but charged for other services. These services included (an optional) hosting service of models, automatic customization of models, and implementation support. The two medical platforms required personal contact and likely also additional services. Sectra explicitly stated that their medical imaging platform can serve as a matchmaker as well as a way to interact (jointly with Sectra and the model provider) to provide a solution based on the AI model that might include services such as training of medical staff and installation.

*Participants charge:* This dimension states how the platform owners charge participants. Contributors could mostly provide offerings for free but had to share their revenue from model consumers with platform owners. Several marketplaces, e.g., Amazon, offered multiple options for charging model contributors, e.g., a model contributor could choose between paying a fixed regular fee and avoiding any costs per transaction or instead paying for each unit sold. On some platforms model contributors could also choose how they wanted to bill customers, which in turn also impacted the revenue flow for the platform.

Some platforms billed for product purchase (one-time) or per usage, i.e., for single access to a model for inference for input or offered free models but charged for accessing them through the rental of their hardware. Large vendors such as Amazon allowed to deploy (purchased) models easily, e.g., almost with the click of a button. For the two platforms offering medical AI products, the billing was not stated. HuggingFace also had a freemium model, charging beyond a usage quota. For OpenAI's GPT store consumers pay a subscription fee for using OpenAI's model, which includes free access to customized models on the GPT stores. Providers are compensated based on the usage of their offerings.

| Arche-type | Universal Marketplace | Specialized Marketplace | Complementary platforms | Open-source exchange platform |
|---|---|---|---|---|
| **Example** | Amazon | Blackford | PyTorch model zoo | PapersWithCode |
| **#Solutions** | 100-10k | <100 | <100-1k | 1k-100k |
| **Value creation** | Centralized storage/ w/o execution | Not stated | Centralized | Centralized /decentralized storage |
| | Internal / User Review | Peer review / Internal | Peer review / Internal | Peer review / Internal / User |
| | No comparison | No comparison | Sometimes | Sometimes |
| | Financial Rewards | Financial | Non-financial/ financial | Non-financial |
| **Value delivery** | API/Download | Indirect (through vendor) | API/web/ Download | Link/Download/API |
| | All domains | Specific | All/ | All |
| | All AI branches | Specific | All | All |
| | All target environments | All | Specific | All |
| | All2All | B2B/ All2All | All2All (with research focus) | All2All with research focus |
| **Value capture** | Sales / Sales of other services | None | Complementary / Sales | None / Sales of other services |

Table 3: Platform archetypes

### 4.2 Platform archetypes

An archetype represents a group of platforms sharing similarities along dimensions with the archetype but also dissimilarities across some dimensions. We distinguish four platform archetypes, i.e., two types of marketplaces for selling and buying AI solutions, complementary platforms, and open-source, research-focused platforms – see Table 3.

We found that universal marketplaces typically aim to attract all AI solutions irrespective of the economic domain, AI branch, and target environment. They commonly provide hosting support for models and various SaaS, particularly to develop models (MLaaS). They are focused on B2B but are, in principle, open to other participants. While market participation might require some form of registration and models might undergo basic testing, entry barriers are fairly low. That is, it is easy to supply a model to a market or to access models, e.g., through standard procedures such as REST APIs. In addition, they often have customer feedback mechanisms and all well-established marketplaces are rather large in terms of their offerings covering multiple domains though smaller ones exist (however, often not for long). In contrast, the two specialized marketplaces were focused on the medical domain. They were very small with internal reviews or referring to external authorities or peer reviewing for model credibility and provided model access only after vendor contact.

Complementary platforms were centralized, rather small, and typically offered state-of-the-art models from academic works for a specific target environment such as Intel's hardware, Pytorch, or OpenAI's infrastructure. Open source research exchange platforms contained the two largest platforms, i.e., hosting 35'000 and 70'000 models. The largest platform (PapersWithCode) actively looked for content by crawling papers and making contributing easy. These platforms did not compensate contributors but also did not charge consumers to access models. HuggingFace offered cloud hosting and a freemium model, i.e., it charged users if they wanted to use their platform beyond a specific quota or were interested in other services, such as MLaaS to identify the best model for a given dataset. However, this did not benefit model providers.

## 5 Discussion

**AI Customization on platforms – Continuous?**

Geske et al. (2021) took the stance that AI service platforms offer customizability on a continuous spectrum, claiming that more development effort implies more customizability. We contrast this viewpoint and provide more nuances. First, we see a major non-continuity depending on whether customization takes place (beyond simple prompt instructions like in the OpenAI GPT store) or not. As soon as customization of AI models is undertaken, a significant amount of expertise on AI technology (and regulation) is needed in many cases - even if no actual development in the form of coding is done, but only data is uploaded and platforms take care of model training or fine-tuning. A company customizing a model must anticipate, for example, how and what to collect data to avoid a mismatch between training data and data used during actual operation. Such a mismatch can lead to potential issues after deployment. Even when customizing leveraging customization services on platforms and using only instructions (as possible on the GPT-store) providing the right instructions and anticipating unintended usages of an AI solution and its consequences is demanding. Second, we believe that speaking of the development of AI as done by Geske et al. (2021) distracts from key AI (model) properties, as AI's functionality is largely determined by data, which must be collected and not developed. That is, an AI might be improved a lot by collecting data and paying experts to label data rather than by developers engaging in any form of coding. Thus, an AI model can be seen as a flexible artifact that can be shaped primarily by data (and instructions for foundation models) rather than by engineering (or coding). We believe that the topic of AI customization in the context of platforms deserves further attention within research, as technical aspects related to programming are (increasingly) less important for the development of AI and customization increasingly takes place on platforms rather than on-premise.

**Evolution of AI exchange platforms – Increasing concentration and company ownership?**

Our work also sheds light on the development of AI exchange platforms over time as we searched for platforms three times with total gaps of about 2.5 years. We found that many platforms initiated by (individual) researchers never reached a considerable AI solution (or user base) and either disappeared or became inactive. While this also holds true for a few commercially oriented platforms, we still found that new commercial platforms emerged, but we found no new platforms developed exclusively by researchers and for researchers. Already large platforms tended to grow further. In particular, HuggingFace has seen tremendous growth in the number of AI solutions jumping from about 70'000 models to more than 350'000 within just two years. Thus, overall, we see a tendency towards market concentration. Still, in the longer-term complementary platforms as well as niche platforms focusing on particular domains (e.g., medical) might withstand the push toward concentration. We also observe a tendency for platforms to increase their service offerings, e.g., more platforms included hosting of models. Furthermore, more exchange platforms also support the testing and creation of AI solutions. The customization process is also becoming increasingly easier and new AI products (such as AI agents) are emerging on platforms. We are not aware of any study which has investigated the evolution of platforms based on comparable empirical evidence. We believe that this is a valuable direction for future research. Especially as looking at the present and the past also allows us to more reliably predict future developments. While it is difficult to foresee the future, the trend of concentration is likely to continue. Furthermore, regulation, high costs for data and models, and very specific requirements might lead to some maintained fragmentation, e.g., as mentioned the healthcare and legal industry could maintain model exchange platforms. One might also envision exchange platforms depending on the risk categories as expressed in the EU AI Act (EU, 2023), i.e., some platforms might not offer models that could be used for higher risk applications as these require different handling by law, e.g., humans might have to double check outcomes.

**Integrating data and AI exchange platforms**

Interestingly, though some platforms provided datasets and models, the two were rather disconnected, e.g., there were limited options to directly import a dataset of the marketplace for model training. On the one hand, this is surprising as the two are strongly interconnected, i.e., AI's behavior is determined by data, and, in turn, certain models are more suitable for some data than for others. Thus, people in search of AI solutions are also quite likely to be in search of data. On the other hand, this is not so surprising as adjustments of AI models are often company (and application) specific and, thus, purchasing data might not be an option but rather application-specific data must be collected. In some cases, an application can be a very narrow concept, e.g., for image recognition changing the image resolution, orientation, scale, and lighting of an object to be recognized could be seen as a different task, i.e., it would require retraining of models to correctly recognize objects given these changes. Additionally, models are commonly pre-trained on large amounts of data, which makes them valuable as training takes time and requires some expertise. If financial (or academic) value is to be gained by offering a trained model on public data on an exchange platform, this is likely taking place as researchers and commercial vendors are eager to capture this value. This claim is aligned with the fact that HuggingFace offers hundreds of thousands of AI models. Today, deep learning models, which constitute the majority and most powerful models, are only based on relatively few fundamentally different technical model architectures(Schneider & Vlachos, 2023). Most differences emerge from training data and minor variations of model architectures, e.g., changing model size to moderate computational resource needs (e.g., for edge device deployment). In conclusion, while we put forth a potential reason why platforms for data and AI solutions are not well-intertwined, we believe that future research is likely to uncover additional aspects as the interrelation appears to be complex.

**Ethics & Explainability?**

Researchers and companies alike might be interested in ensuring the ethical use of their AI solutions. However, existing platforms provide little support in doing so. Most models were not assessed or discussed with respect to common issues of AI, e.g., biases, fairness, robustness, and security. This is somewhat surprising, given that academia extensively treats these topics, and ethical debates surrounding AI are increasingly in the news. There were a few exceptions, e.g., on HuggingFace we could find the following statement "We encourage to check GPT2[Link] to know more about usage, limitations, and potential biases." Alsom on HuggingFace, models were encompanied with model cards providing information on the model that occasionally contained intended and out-of-scope uses. OpenAI had a more elaborate content policy aimed at mitigating the risks of offering unethical customized AI models.

Furthermore, while explainability is a key topic in academia (Meske et al., 2022; Schneider, 2024), it is not prevalent (as an additional service) on most AI exchange platforms. This could be due to the fact that it is a very rapidly evolving area, where there is still a need to develop better methods. Either way, (functionality for) model understanding could foster trust, which is an important factor for platforms (Verhagen et al., 2006).

We believe that researchers should contribute more to better understand these phenomena.

**Leveraging communities as external developers**

Successful platforms "invert their firm using external developers" (Parker et al., 2017). This might hold even more true given the scarcity and high costs of AI experts. Still, it is not clear how to best manage communities, in particular, as AI exchange platforms add researchers (and hobbyists) to a set of contributors. The TensorFlow model garden (by Google) provides an interesting case, as it evolved in how it manages contributions, i.e., it created separate branches for models managed by researchers, the community, and "official" models. Prior, there was only the "official" branch, which was used by all users. This separation is an interesting idea, as (i) it reduces reputation risks in case models are flawed (as the research models are officially managed by researchers), (ii) it reduces costs of oversight (as errors are not so critical – similar to an alpha version of software), and (iii) it allows to attract more contributions due to less stringent contribution procedures. However, we encourage researchers to elaborate on the alleged reasons in separate studies. Furthermore, open source models have enjoyed a lot of attention, leading to countless hours being invested in creating fine-tuned models for specific tasks or in compressing such models to make them executable on ordinary PCs (without expensive GPUs). For example, the Meta's Llama-3 model has more than 10k model variants on Huggingface and Meta officially announced that open sourcing AI is the way forward (Meta, 2024). In contrast, commercial model providers like OpenAI allow to share knowledge enhanced versions of models in a much simpler way, e.g., with a few clicks one can create and share a custom GPT (OpenAI, 2023). Technically, these custom GPT models do not differ from the original model, but they execute a crafted user prompt and can possibly access specific user knowledge. These two models also attract different communities, e.g., technical AI experts in case of Meta and non-technical (domain experts) in the case of OpenAI. Clearly, this raises questions about what is the better model or is there a way to leverage both communities?

**Research 2 Business (R2B) and the relevance of research policies**

The magnitude of knowledge transfer and speed from research to industry in the field of AI can be said to be exceptional in sciences, with tens of thousands of models being contributed concurrently with publications. Submissions and publication guidelines (which we denote as research policy) can have a strong practical impact that goes well beyond their intended consequences. That is, while academic outlets' primary reason is to ensure reproducibility and transparency (Burton-Jones et al., 2021; Gundersen &

Kjensmo, 2018; Pineau et al., 2021), these research policies also have positive side effects on technology transfer, adoption, ethics and transparency in practice - not least, since many large industrial research institutions and maybe more so their researchers (e.g., from OpenAI, Google, Meta) aim at publishing at academic conferences. For example, AAAI, one of the largest AI conferences, has a reproducibility checklist, where researchers are asked to state if "All source code required for conducting experiments will be made publicly available upon publication of the paper with a license that allows free usage for research purposes."[8] Other disciplines like information systems might also benefit from adopting such policies.

Overall, we believe there are multiple possible reasons for the widespread adoption of AI outputs from research: First, conferences and journals increasingly push researchers to not only publish ideas, and theories, paired with explanations and evaluation results, but also offer an implementation of their ideas in the form of code. Second, implementations by researchers are easily usable by businesses. They are mostly developed with frameworks also used in industry, such as TensorFlow and Pytorch. Rather than re-implementing ideas from papers which is difficult, companies (and fellow researchers) can easily access repositories providing access to ready-to-use models. Put differently, researchers and practitioners speak the same language. Both researchers and industry practitioners also often share the same metrics, e.g., model accuracy. Third, AI models are easy to adopt since they are straightforward to adjust (using problem-specific data) and require little code. Fourth, AI models require little effort to share to a wide audience thanks to AI exchange platforms and public repositories such as GitHub. Fifth, for adoption by businesses, it is important that there are no license restrictions. While current policies of journals and conferences do not necessarily enforce this, i.e., they generally only ask for code release for research purposes, many researchers still grant unrestricted usage. There are also additional factors contributing to this growth, such as an increase in AI funding (and thus research outputs) or an increased belief that code helps in citation count (as shown for releasing research data (Colavizza et al., 2020)).

**Limitations**

Many AI exchange platforms appear quickly but also vanish. As such, our research is a snapshot in time. This implies that the proposed taxonomy is not static but will likely evolve over time. Additionally, many platforms lacked information on their websites that could have been interesting for our study, e.g., details on internal model review policies. Although the taxonomy is built upon the analysis of actual platforms, the data collection itself is subject to interpretation. Other researchers might derive other dimensions and characteristics. Regarding the selection of the platforms, there is a continuum of offerings on platforms covering exchanging only (raw) AI models onto software, where AI plays a small piece and plays a rather irrelevant role. Our requirement that AI is central includes just one end of the spectrum.

---

[8] https://aaai.org/Conferences/AAAI-22/reproducibility-checklist/

# 6 Conclusions

AI is fast-paced and transformational. As customization of AI becomes easier (as witnessed, e.g., by OpenAI's GPT store), AI models are likely to be found in a growing number of products fostering the need to facilitate the exchange of these models. At the same time, such platforms are in a unique position to capture value through economy of scope and implement policies to foster ethical usage of AI. While we see some platforms moving in this direction, existing services of these platforms differ largely and are likely to evolve further. Our taxonomy is thus a snapshot in time that helps to understand the offerings of platforms matching the needs and preferences of both solution providers and consumers. Our analysis revealed also that AI exchange platforms constitute a positive example for facilitating transfer from research to industry on a large scale – not at least thanks to research policies by conferences and journals.

# 7 Statements and Declarations

**Funding**

None/Not applicable

**Conflicts of interest/Competing interests**

None/ Not applicable

**Data availability**

Data is provided within the manuscript.

**Author contributions**

R.A. supported conceptualization, writing and provided reviews with feedback; the rest was done by the first author J.S.

# References

Bakos, J. Y. (1991). A strategic analysis of electronic marketplaces. *MIS Quarterly*, 295–310.

Basalla, M., Schneider, J., & vom Brocke, J. (2022). Creativity of deep learning: Conceptualization and assessment. *Proceedings of the 14th International Conference on Agents and Artificial Intelligence*.

Berente, N., Gu, B., Recker, J., & Santhanam, R. (2021). Managing artificial intelligence. *MIS Quarterly*, *45*(3).

Blaschke, M., Haki, K., Aier, S., & Winter, R. (2019). Taxonomy of digital platforms: A platform architecture perspective. *Wirtschaftsinformatik*.

Bock, A. C., & Frank, U. (2021). Low-code platform. *Business & Information Systems Engineering*, *63*, 733–740.

Bommasani, R., Hudson, D. A., Adeli, E., Altman, R., Arora, S., von Arx, S., Bernstein, M. S., Bohg, J., Bosselut, A., Brunskill, E., & others. (2021). On the opportunities and risks of foundation models. *arXiv Preprint arXiv:2108.07258*.

Burton-Jones, A., Oborn, E., Padmanabhan, B., Boh, W. F., & Kohli, R. (2021). *Research Exchange-Sept. 13, 2021" Advancing Research Transparency-A View from MISQ"*.

Chakraborty, A., Alam, M., Dey, V., Chattopadhyay, A., & Mukhopadhyay, D. (2018). Adversarial attacks and defences: A survey. *arXiv Preprint arXiv:1810.00069*.

Colavizza, G., Hrynaszkiewicz, I., Staden, I., Whitaker, K., & McGillivray, B. (2020). The citation advantage of linking publications to research data. *PloS One*, *15*(4), e0230416.

De Reuver, M., Sørensen, C., & Basole, R. C. (2018). The digital platform: A research agenda. *Journal of Information Technology*, *33*(2), 124–135.

Dettmers, T., Pagnoni, A., Holtzman, A., & Zettlemoyer, L. (2023). Qlora: Efficient finetuning of quantized llms. *arXiv Preprint arXiv:2305.14314*.

EU. (2023). *EU AI Act*. https://artificialintelligenceact.eu/

Fang, M. (2024, August 16). *A Deep Dive Into IDC's Global AI and Generative AI Spending*. https://blogs.idc.com/2024/08/16/a-deep-dive-into-idcs-global-ai-and-generative-ai-spending/

Freichel, C., Fieger, J., & Winkelmann, A. (2021). Developing a Taxonomy for Digital Platforms– A Conceptual Approach. *Hawaii International Conference on System Sciences*.

Gawer, A. (2014). Bridging differing perspectives on technological platforms: Toward an integrative framework. *Research Policy*, *43*(7), 1239–1249.

Geske, F., Hofmann, P., Lämmermann, L., Schlatt, V., & Urbach, N. (2021). Gateways to Artificial Intelligence: Developing a Taxonomy for AI Service Platforms. *ECIS 2021 Research Papers*.

Gorwa, R., & Veale, M. (2023). Moderating Model Marketplaces: Platform Governance Puzzles for AI Intermediaries. *arXiv Preprint arXiv:2311.12573*.

Grewal, D., Guha, A., Satornino, C. B., & Schweiger, E. B. (2021). Artificial intelligence: The light and the darkness. *Journal of Business Research*, *136*, 229–236.

Gundersen, O. E., & Kjensmo, S. (2018). State of the art: Reproducibility in artificial intelligence. *Proceedings of the AAAI Conference on Artificial Intelligence*, *32*(1).

Jöhnk, J., Weißert, M., & Wyrtki, K. (2021). Ready or not, AI comes—An interview study of organizational AI readiness factors. *Business & Information Systems Engineering*, *63*(1), 5–20.

Kumar, A., Finley, B., Braud, T., Tarkoma, S., & Hui, P. (2021). Sketching an ai marketplace: Tech, economic, and regulatory aspects. *IEEE Access*, *9*, 13761–13774.

Kurtulmus, A. B., & Daniel, K. (2018). Trustless machine learning contracts; evaluating and exchanging machine learning models on the ethereum blockchain. *arXiv Preprint arXiv:1802.10185*.

Lins, S., Pandl, K. D., Teigeler, H., Thiebes, S., Bayer, C., & Sunyaev, A. (2021). Artificial Intelligence as a Service. *Business & Information Systems Engineering*, *63*(4), 441–456.

Lyytinen, K., Yoo, Y., & Boland Jr, R. J. (2016). Digital product innovation within four classes of innovation networks. *Information Systems Journal*, *26*(1), 47–75.

Meske, C., Bunde, E., Schneider, J., & Gersch, M. (2022). Explainable artificial intelligence: Objectives, stakeholders, and future research opportunities. *Information Systems Management*, *39*(1), 53–63.

Meta. (2024, July). *Open Source AI is the Path Forward*. https://about.fb.com/news/2024/07/open-source-ai-is-the-path-forward/

Movahedi, B. M., Lavassani, K. M., & Kumar, V. (2012). E-marketplace emergence: Evolution, developments and classification. *Journal of Electronic Commerce in Organizations (JECO)*, *10*(1), 14–32.

Nickerson, R. C., Varshney, U., & Muntermann, J. (2013). A method for taxonomy development and its application in information systems. *European Journal of Information Systems*, *22*(3), 336–359.

OpenAI. (2023). *Introducing the GPT Store*. https://openai.com/blog/introducing-the-gpt-store

Parker, G., Van Alstyne, M., & Jiang, X. (2017). Platform ecosystems. *Mis Quarterly*, *41*(1), 255–266.

Pavlou, P. A., & Dimoka, A. (2006). The nature and role of feedback text comments in online marketplaces: Implications for trust building, price premiums, and seller differentiation. *Information Systems Research*, *17*(4), 392–414.

Pei, J., Fernandez, R. C., & Yu, X. (2023). Data and ai model markets: Opportunities for data and model sharing, discovery, and integration. *Proceedings of the VLDB Endowment*, *16*(12), 3872–3873.

Pineau, J., Vincent-Lamarre, P., Sinha, K., Larivière, V., Beygelzimer, A., d'Alché-Buc, F., Fox, E., & Larochelle, H. (2021). Improving reproducibility in machine learning research: A report from the NeurIPS 2019 reproducibility program. *Journal of Machine Learning Research*, *22*.

Qian, M., Musa, A. A., Biswas, M., Guo, Y., Liao, W., & Yu, W. (2025). Survey of Artificial Intelligence Model Marketplace. *Future Internet*, *17*(1), 35. https://doi.org/10.3390/fi17010035

Rai, A., Constantinides, P., & Sarker, S. (2019). Next Generation Digital Platforms: Toward Human-AI Hybrids. *Mis Quarterly*, *43*(1), iii–ix.

Remane, G., Hanelt, A., Tesch, J. F., & Kolbe, L. M. (2017). The business model pattern database—A tool for systematic business model innovation. *International Journal of Innovation Management*, *21*(01), 1750004.

Sarpatwar, K., Sitaramagiridharganesh Ganapavarapu, V., Shanmugam, K., Rahman, A., & Vaculin, R. (2019). Blockchain enabled AI marketplace: The price you pay for trust. *Proceedings of the IEEE/CVF Conference on Computer Vision and Pattern Recognition Workshops*, 0–0.

Schneider, J. (2024). Explainable Generative AI (GenXAI): A Survey, Conceptualization, and Research Agenda. *Artificial Intelligence Review*.

Schneider, J. (2025). Generative to Agentic AI: Survey, Conceptualization, and Challenges. *arXiv Preprint arXiv:2504.18875*. https://arxiv.org/abs/2504.18875

Schneider, J., Meske, C., & Kuss, P. (2024). Foundation Models. *Business Information Systems Engineering*.

Schneider, J., & Vlachos, M. (2023). A Survey of Deep Learning: From Activations to Transformers. *arXiv Preprint arXiv:2302.00722*.

Schneider, J., Vlachos, M., & Meske, C. (2022). Deceptive AI explanations: Creation and detection. *International Conference on Agents and Artificial Intelligence (ICAART)*.

Somy, N. B., Kannan, K., Arya, V., Hans, S., Singh, A., Lohia, P., & Mehta, S. (2019). Ownership preserving AI market places using blockchain. *2019 IEEE International Conference on Blockchain (Blockchain)*, 156–165.

Täuscher, K., & Laudien, S. M. (2018). Understanding platform business models: A mixed methods study of marketplaces. *European Management Journal*, *36*(3), 319–329.

Truex, S., Liu, L., Gursoy, M. E., Yu, L., & Wei, W. (2019). Demystifying membership inference attacks in machine learning as a service. *IEEE Transactions on Services Computing*.

Van den Broek, E., Sergeeva, A., & Huysman, M. (2021). When the Machine Meets the Expert: An Ethnography of Developing AI for Hiring. *MIS Quarterly*, *45*(3).

Verhagen, T., Meents, S., & Tan, Y.-H. (2006). Perceived risk and trust associated with purchasing at electronic marketplaces. *European Journal of Information Systems*, *15*, 542–555.

Williams, K., Chatterjee, S., & Rossi, M. (2008). Design of emerging digital services: A taxonomy. *European Journal of Information Systems*, *17*(5), 505–517.

Wolpert, D. H., & Macready, W. G. (1997). No free lunch theorems for optimization. *IEEE Transactions on Evolutionary Computation*, *1*(1), 67–82.

Xiu, M., Jiang, Z. M. J., & Adams, B. (2020). An Exploratory Study of Machine Learning Model Stores. *IEEE Software*, *38*(1), 114–122.

Xu, M., Liu, J., Liu, Y., Lin, F. X., Liu, Y., & Liu, X. (2019). A first look at deep learning apps on smartphones. *The World Wide Web Conference*, 2125–2136.

Yao, Y., Xiao, Z., Wang, B., Viswanath, B., Zheng, H., & Zhao, B. Y. (2017). Complexity vs. Performance: Empirical analysis of machine learning as a service. *Proceedings of the 2017 Internet Measurement Conference*, 384–397.

Yin, R. K. (2009). *Case study research: Design and methods* (Vol. 5). sage.